\documentclass[useAMS,usenatbib,usegraphicx]{mn2e} %
\usepackage[latin9]{inputenc}
\usepackage{amsmath}
\usepackage{amssymb}
\usepackage{natbib}
\begin{document}

\newcommand{\avg}[1]{\left\langle #1 \right\rangle}
\newcommand{\mc}[1]{\mathcal{#1}}
\newcommand{\mb}[1]{\mathbf{#1}}
\newcommand{\tmb}[1]{\tilde{\mathbf{#1}}}
\newcommand{\scal}[2]{\mathopen ( #1 . #2 \mathclose )}
\newcommand{\pv}[2]{#1 \wedge #2}
\newcommand{\modif}[1]{#1} %{[MODIFIED] #1 [/MODIFIED]}
\newcommand{\ab}{a_1}
\newcommand{\eb}{e_1}
\newcommand{\Mb}{M_1}
\newcommand{\MMb}{\mathcal{M}_{01}}
\newcommand{\nub}{\beta_1}
\newcommand{\vrb}{\mb{r}_1}
\newcommand{\rb}{r_1}

\newcommand{\rev}[1]{#1}

%\spnewtheorem*{myremark}{Remark:}{\bf}{\rm}

\numberwithin{equation}{section}

\title{High inclination orbits in the secular quadrupolar three-body problem}
\author[F. Farago and J. Laskar]{F. Farago\thanks{E-mail: farago@imcce.fr; laskar@imcce.fr} and J. Laskar\footnotemark[1]\\
ASD, IMCCE-CNRS UMR8028, Observatoire de Paris, UPMC,
           77 avenue Denfert-Rochereau, 75014 Paris, France}

\date{\today}

\maketitle

\begin{abstract} The Lidov-Kozai mechanism \citep{1962AJ.....67..591K,1962P&SS....9..719L} allows a body
to periodically exchange its eccentricity with inclination. It was first
discussed in the framework of the quadrupolar secular restricted three-body problem, where the massless particle is
the inner body, and later extended to the quadrupolar secular nonrestricted three body problem
\citep{1969CeMec...1..200H,1976CeMec..13..471L,1994CeMDA..58..245F}. In this paper, we propose a different point
of view on the problem by looking first at the restricted problem where the massless particle is the outer
body. In this situation, equilibria at high mutual inclination appear \rev{\citep{2006PhyD..213...15P}}, which correspond to the population
of stable particles that \citet{2008MNRAS.390.1377V,2009MNRAS.394.1721V} find in stable, high inclination
circumbinary orbits around one of the components of the quadruple star HD 98800. We provide \rev{a simple analytical
framework using a vectorial formalism for these situations. We also look} at the evolution of these high inclination equilibria in the
non restricted case. \end{abstract}


\section{Introduction}\label{se:intro}

As it is known, the secular three-body problem after node reduction has two degrees of freedom (e. g.
\cite{1905QB351.P73......,2002CeMDA..84..283M}). However, due to what \citet{1976CeMec..13..471L} called a
{\em happy coincidence}, this problem is integrable when it is expanded up to order two in the ratio of
semi-major axes, i.e. at the quadrupolar approximation. Indeed, the argument of perihelion of the outer body
does not explicitly appear in the quadrupolar expansion of the secular problem, thus giving one more integral
of motion linked to the eccentricity of the outer body.

The limiting case where the inner body has no mass has been extensively studied
\citep{1962AJ.....67..591K,1962P&SS....9..719L,2007CeMDA..98...67K}. We will call this problem the {\em inner
restricted problem}, while the converse case where the two inner bodies are massive and the outer body is
massless will be called the {\em outer restricted problem}. In the inner restricted case, the conservation of the
normal component of the angular momentum enables the inner particle to periodically exchange its eccentricity
with inclination (the so-called Lidov-Kozai mechanism). The inner
restricted model is well suited when the inner body has a small mass with respect to the other two. However,
when looking at higher mass ratios, for example in triple star systems, this is no longer justified.

Since the Hamiltonian of the quadrupolar problem of three masses is very similar to that of the inner
restricted problem when it is written in elliptic variables, the study of the massive problem has mainly
focused on the dynamics of the two inner bodies
\citep{1969CeMec...1..200H,1976CeMec..13..471L,1994CeMDA..58..245F}. These previous works completely
classified the different dynamical regimes and bifurcations, using the equations of motion of the inner binary.

There is however another limit-case to the massive problem, which is the outer restricted problem.
\rev{\citet{2006PhyD..213...15P} have studied this case and discussed the existence and stability
of equilibria in the non-averaged system using the framework of KAM theory. We give here a very simple
model of the outer restricted case which provides an alternate formulation of these previous results and
which is directly usable in an astronomical context. We also fully describe the possible motions of the bodies
and give an analytical expression of their frequencies.} We use this model to explain the results of
\citet{2008MNRAS.390.1377V,2009MNRAS.394.1721V}, who find populations of particles at very high inclinations
around one of the components of the double-binary star HD 98800, which are stable even under the perturbation
of the other component. We then look at the quadrupolar problem of three masses from the perspective of the
outer restricted problem \rev{and show how the inner and outer restricted cases are related to the general case}.
Vectorial methods as developed by \citet{2006Icar..185..312B,2009Icar..201..750B,2009AJ....137.3706T} are extremely
well suited for this approach.

\section{Secular outer restricted problem}\label{se:rtbp}
\subsection{Derivation of the Hamiltonian}

We consider here the case of a massless particle orbiting a central binary object. We do not restrict
ourselves with respect to inclinations or eccentricities. The components of the binary have masses $m_0$ and
$m_1$, the binary's total mass is $\MMb=m_0+m_1$ and its reduced mass is $\nub=m_0m_1/(m_0+m_1)$. The two
massive components have barycentric positions $\mb{u}_0\ \text{and}\ \mb{u}_1$. We
also denote $\delta=m_0/\MMb$ and $\vrb=\mb{u}_1-\mb{u}_0$, and $\mb{r}_2$ is the position of the outer particle
relatively to the barycentre of the inner binary. Using these notations, the particle has the
following Hamiltonian:

\begin{equation}
 H = \frac{\tilde{\mb{r}}^2_2}{2}-G\left(\frac{m_0}{|\mb{r}_2-\mb{u}_0|}+\frac{m_1}{|\mb{r}_2-\mb{u}_1|}\right)\ ,
\end{equation}

\noindent where $\tilde{\mb{r}}_2=\dot{\mb{r}}_2$ is the canonical momentum associated to the position of the
massless particle, $\mb{r}_2$. Since $\mb{u}_0=-(1-\delta)\vrb$ and $\mb{u}_1=\delta\vrb$, we can rewrite the
Hamiltonian as:

\begin{equation}
 H=\frac{\tilde{\mb{r}}^2_2}{2}-G\left(\frac{m_0}{|\mb{r}_2+(1-\delta)\vrb|}
 +\frac{m_1}{|\mb{r}_2-\delta\vrb|}\right)\ .
\end{equation}

We now suppose that $\rb \ll r_2$ and expand the Hamiltonian to order 2 in $r_1/r_2$:

\begin{equation}\label{eq:hamqp}
 H = \left(\frac{\tilde{\mb{r}}^2_2}{2}-\frac{G\MMb}{r_2}\right)
 -\frac{G\nub}{2r_2^3}\left(3\frac{\scal{\mb{r}_2}{\vrb}^2}{r_2^2}-\rb^2\right)\ .
\end{equation}

The first term is the Keplerian energy of the particle interacting with the binary, seen as a point mass
$\MMb$. It is equal to $-G\MMb/2a_2$, where $a_2$ is the semi major axis of the particle.

Since we are interested in the secular behaviour of the particle, we average this quadrupolar Hamiltonian
over the mean anomalies of the binary ($\Mb$) and of the particle ($M_2$). In order to do this, we first
introduce four unit vectors: $(\mb{i},\mb{j},\mb{k})$ are bound to the orbit of the binary, remain constant,
and will provide a natural reference frame; $\mb{w}$ is bound to the orbit of the particle and will vary.
More precisely, $\mb{i}$ points in the direction of the perihelion of the binary, $\mb{k}$ is colinear to the
angular momentum of the binary, and $\mb{j}=\pv{\mb{k}}{\mb{i}}$; the last vector $\mb{w}$ is colinear to the
angular momentum of the massless particle.

We can then compute the following averaged quantities, where quantities indexed with $1$ relate to the
binary, quantities with index $2$ relate to the particle, and $\mb{u}$ is an arbitrary fixed vector (see for
instance the appendix of \citet{2006Icar..185..312B}):

\begin{eqnarray}
 \avg{\rb^2}_{\Mb} & = & \ab^2\left(1+\frac{3}{2}\eb^2\right)\ ; \\
 \avg{\scal{\vrb}{\mb{r}_2}^2}_{\Mb} & = & \frac{\ab^2}{2}(r_2^2-\scal{\mb{k}}{\mb{r}_2}^2)\nonumber\\
    &&+\frac{\ab^2\eb^2}{2}(4\scal{\mb{i}}{\mb{r}_2}^2 - \scal{\mb{j}}{\mb{r}_2}^2)\ ; \\
 \avg{\frac{1}{r_2^3}}_{M_2} & = & \frac{1}{a_2^3(1-e_2^2)^{3/2}}\ ; \\
 \avg{\frac{\scal{\mb{r}_2}{\mb{u}}^2}{r_2^5}}_{M_2} & = & 
    \frac{u^2-\scal{\mb{w}}{\mb{u}}^2}{2a_2^3(1-e_2^2)^{3/2}}\ .
\end{eqnarray}

The substitution of these expressions in \eqref{eq:hamqp} yields:

\begin{multline}\label{eq:hmoy}
\avg{H}_{\Mb,M_2} = -\frac{G\MMb}{2a_2} -\frac{3}{8}\frac{G\nub\ab^2}{a_2^3(1-e_2^2)^{3/2}}\times\\
\left[\left(\eb^2-\frac{1}{3}\right)
+\scal{\mb{k}}{\mb{w}}^2-\eb^2(4\scal{\mb{i}}{\mb{w}}^2-\scal{\mb{j}}{\mb{w}}^2)\right]
\end{multline}

Since the particle has no mass, the only variable element of the binary is its mean anomaly $\Mb$. After
averaging over this angle, it is no longer present in the Hamiltonian. After averaging over the mean anomaly
of the particle, its semi major axis $a_2$ becomes constant. Moreover, $\mb{w}=\sin i_2 \sin\Omega_2\ \mb{i} -
\sin i_2 \cos\Omega_2\ \mb{j} + \cos i_2\ \mb{k},$ so the argument of pericentre $\omega_2$ of the particle
does not appear in the averaged Hamiltonian. Hence, at the quadrupolar order, the conjugate momentum
associated to $\omega_2$, i.e. the norm of the angular momentum of the particle $G_2=\sqrt{G\MMb
a_2(1-e_2^2)}$, is constant. Therefore the eccentricity $e_2$ of the particle is constant. This fact is a
feature of the quadrupolar expansion, not a property of the restricted problem. As such it remains true when
the outer body has a non-zero mass (see section \ref{se:ftbp}). This is the {\em happy coincidence} that
\citet{1976CeMec..13..471L} noted. Finally, only one degree of freedom remains, related to the couple
$(i_2,\Omega_2)$.

If we drop the constant terms in \eqref{eq:hmoy}, and introduce the mean motion $n_1$ of the binary into
the Hamiltonian ($n_1^2\ab^3=G\MMb$), we get the following expression\footnote{We will from now write
$\avg{H}$ for the averaged Hamiltonian, omitting the subscripts $M_1,M_2$.} \rev{(see also eq. 10 in
\citep{2006PhyD..213...15P})}:

% \begin{equation}
%  \avg{H}_{\Mb,M} = -\frac{3}{8}\frac{G\nub\ab^2}{a^3(1-e^2)^{3/2}}
%  \left[\scal{\mb{k}}{\mb{w}}^2-\eb^2(4\scal{\mb{i}}{\mb{w}}^2-\scal{\mb{j}}{\mb{w}}^2) \right]
% \end{equation}

\begin{equation}
\avg{H} = -\frac{\alpha G_2}{2}
\left[\scal{\mb{k}}{\mb{w}}^2-\eb^2(4\scal{\mb{i}}{\mb{w}}^2-\scal{\mb{j}}{\mb{w}}^2) \right]\ ,
\end{equation}
where
\begin{equation}
\alpha = \frac{3}{4}n_1 \left(\frac{\ab}{a_2}\right)^{7/2} \frac{\nub}{\MMb}\frac{1}{(1-e_2^2)^2}\ .
\end{equation}

This Hamiltonian can be rewritten in a very compact form as:
\begin{equation}
 \avg{H} = -\frac{1}{2}\phantom{.}^t\mb{w}.\mb{T}.\mb{w}\ ,\ \text{where:}
\end{equation}
\begin{equation}
\mb{T} = \alpha G_2\left(\begin{array}{ccc}
                -4\eb^2 & 0 & 0 \\
                0 & \eb^2 & 0 \\
                0 & 0 & 1
               \end{array}\right)\ .
\end{equation}

We also give the expression of the Hamiltonian using the inclination and the node of the particle:

\begin{equation}\label{eq:HrestIO}
 \avg{H}=-\frac{\alpha G_2}{4}\left[2\cos^2i_2 - \eb^2\sin^2i_2\left(3-5\cos2\Omega_2\right)\right]\ .
\end{equation}

\subsection{Equations of motion}\label{se:restreqns}

As discussed in \citep{2006Icar..185..312B}, the equations of motion for $\mb{w}$ are simply 
obtained by:

\begin{equation}
 \dot{\mb{w}} = \frac{1}{G_2}\nabla_{\mb{w}}\avg{H}\wedge\mb{w}\ .
\end{equation}

After computing the gradient, we find:

\begin{equation}
 \dot{\mb{w}} = -\alpha
  \left[(\mathbf{k}.\mathbf{w})(\mathbf{k}\wedge\mathbf{w})
       -\eb^2(4(\mathbf{i}.\mathbf{w})(\mathbf{i}\wedge\mathbf{w})
              -(\mathbf{j}.\mathbf{w})(\mathbf{j}\wedge\mathbf{w}))\right]\ .
\end{equation}

If we note $x=(\mathbf{i}.\mathbf{w})$, $y=(\mathbf{j}.\mathbf{w})$, and $z=(\mathbf{k}.\mathbf{w})$,
we get the following system for $(x,y,z)$:

\begin{eqnarray}
\dot{x} & = & \alpha(1-\eb^2)yz\ ; \label{eq:xdot}\\
\dot{y} & = & -\alpha(1+4\eb^2)zx\ ; \\
\dot{z} & = & 5\alpha \eb^2 xy\ .\label{eq:zdot}
\end{eqnarray}

In these variables, the fact that $\mb{w}$ is a unit vector and the energy integral translate into the
following equalities:

\begin{eqnarray}
 x^2+y^2+z^2 & = & 1\ ; \label{eq:wsphere}\\
 z^2 - \eb^2(4x^2-y^2) & = & h = \text{Cst}\ . \label{eq:Hsurf}
\end{eqnarray}

The system of three differential equations \eqref{eq:xdot}--\eqref{eq:zdot} has thus two independent first
integrals and is as such integrable. It is also straightforward from these two relations that 
\begin{equation}\label{eq:hineg}
	-4\eb^2\leqslant h\leqslant 1\ .
\end{equation}

\subsection{Motion of a massless body around a circular binary}

In the case of a circular binary, the Hamiltonian and the equations of motion greatly simplify\footnote{
\rev{The next non-zero term of the Hamiltonian which is the fourth order in $(a_1/a_2)$ plays
an important part in the circular case as has been discussed in detail by \citet{2006CeMDA..95...81P}.}}:

\begin{equation}
 \avg{H} = -\frac{\alpha G_2}{2}\scal{\mb{k}}{\mb{w}}^2\ ,
\end{equation}
\begin{equation}
  \dot{\mb{w}} = -\alpha
  (\mathbf{k}.\mathbf{w})(\mathbf{k}\wedge\mathbf{w})
\end{equation}

The scalar product $(\mathbf{k}.\mathbf{w})=\cos i_2$ remains constant, and the nodes of the orbit of the
particle simply precess around the angular momentum of the binary, with a constant precession rate:
\begin{equation}
 \dot{\Omega}=-\alpha\cos i_2 =
 -\frac{3}{4}n_1 \left(\frac{\ab}{a_2}\right)^{7/2} \frac{\nub}{\MMb}\frac{\cos i_2}{(1-e_2^2)^2}
\end{equation}

This precession is equivalent to the precession generated by the quadrupolar potential of a circular and
homogeneous ring of mass $\beta_1$ and of radius $a_1$ \rev{following an idea which can be traced back to
Gauss (see \citep{2009MNRAS.394.1085T} and references therein)}.

\subsection{Motion of a massless body around an elliptic binary}
\subsubsection{Qualitative overview}

When the binary is elliptic, the situation changes and cannot be explained any longer by the quadrupolar torque
of a circular ring. If we substitute $z^2$ in \eqref{eq:Hsurf} using \eqref{eq:wsphere}, we get:
\begin{align}\label{eq:hinterw}
 (1+4\eb^2)x^2+(1-\eb^2)y^2 &=1-h\geqslant 0\ ,\\
  x^2+y^2+z^2 &= 1\ .
\end{align}

\begin{figure} 
    \includegraphics[width=\columnwidth]{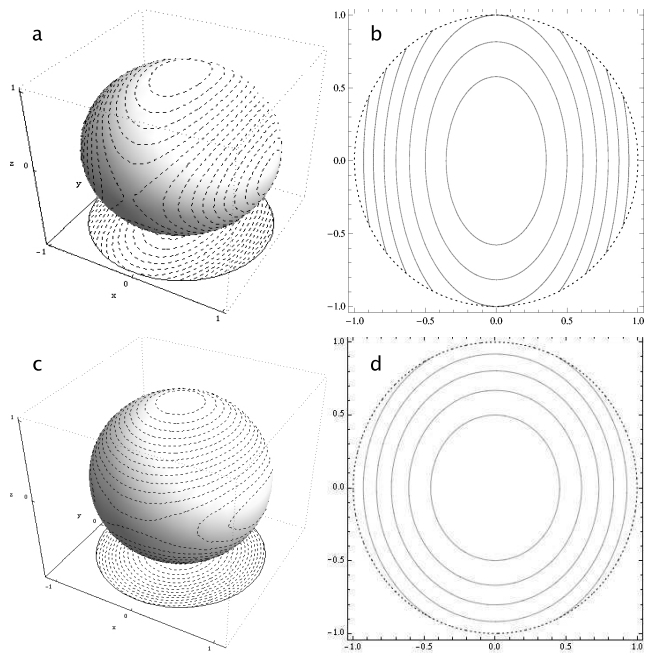} 
    \caption{\label{fig:Hxy}Intersections of the
    energy surfaces and the unit angular momentum sphere (a) and its projection in the $(x,y)$ plane (b) for
    $\eb=0.5$. Intersections of the energy surfaces and the unit angular momentum sphere (c) and its 
    projection in the $(x,y)$ plane (d) for $\eb=0.2$.} 
\end{figure}

The intersections of the energy surfaces and the normalized angular momentum sphere of the particle can thus be
seen as the intersections of elliptic cylinders with the unit sphere. For a given value of the energy $h$, the
extremity of the unit angular momentum vector of the particle $\mb{w}$ will move on the intersection of the
corresponding cylinder with the unit sphere. Figures \ref{fig:Hxy}.a and c show these intersections for
different values of the energy as dotted lines drawn on the unit sphere, in two situations where the binary has
an eccentricity of $0.5$ and $0.2$ respectively. The three axes correspond to the scalar products $x$, $y$ and
$z$ that are defined in section \ref{se:restreqns}.

There are four visible kinds of trajectories: closed trajectories around the two poles of the sphere
$(x,y,z)=(0,0,\pm 1)$, and closed trajectories around the points $(x,y,z)=(\pm 1,0,0)$.

When the extremity of the angular momentum of the particle $\mb{w}$ follows a trajectory around the north
pole, it means that it precesses around the angular momentum of the binary $\mb{k}$ with an inclination that
is strictly inferior to $90^\circ$: in this case, the orbital motion of the particle is prograde relatively
to the orbital motion of the binary.

When the extremity of the angular momentum of the particle $\mb{w}$ follows a trajectory around the south
pole, it means that it precesses around the opposite of the angular momentum of the binary, $-\mb{k}$, with
an inclination that is strictly superior to $90^\circ$: in this case, the orbital motion of the particle is
retrograde relatively to the orbital motion of the binary.

When the extremity of the angular momentum of the particle $\mb{w}$ follows a trajectory around one of the
two points $(x,y,z)=(\pm 1,0,0)$, it precesses around the direction of the perihelion of the binary or the
opposite of this direction. In this case, the inclination oscillates around $\pm 90^\circ$.

\subsubsection{Frequencies} 

The frequencies of these motions can be found analytically. Indeed, using equation \eqref{eq:hinterw}, we see
that $x$ and $y$ are on ellipses or arcs of ellipses bounded by the unit circle (figures \ref{fig:Hxy}.b and d
show respectively the cases where $\eb=0.5$ and $0.2$). Thus, there is an angle $\phi$ such that:

\begin{eqnarray}
 x & = & \sqrt{\frac{1-h}{1+4\eb^2}} \cos\phi\ , \label{eq:xell}\\
 y & = & \sqrt{\frac{1-h}{1-\eb^2}} \sin \phi\ .
\end{eqnarray}

Using \eqref{eq:wsphere}, we can then express $z^2$ as:

\begin{equation}\label{eq:z2phi}
 z^2 = \frac{h+4\eb^2}{1+4\eb^2} - \frac{5\eb^2}{1+4\eb^2}\frac{1-h}{1-\eb^2} \sin^2\phi\ .
\end{equation}

There are two opposite values of $z$ for each $\phi$, reflecting the symmetry of the system with respect to
the orbital plane of the binary. If we use expression \eqref{eq:z2phi} in combination with equation
\eqref{eq:zdot}, we obtain the following equation for $\dot{\phi}$:

\begin{eqnarray}
 \dot{\phi}&=&\mp\alpha\sqrt{(1-\eb^2)(h+4\eb^2)}
\sqrt{1-k^2\sin^2\phi}\ , \label{eq:dotphi}\\\
k^2&=&\frac{5\eb^2}{1-\eb^2}\frac{1-h}{h+4\eb^2}\ .
\end{eqnarray}

The constant $k^2$ is positive because of relation \eqref{eq:hineg}. The value $k^2=1$ defines a
limit between two dynamical regimes. If $k^2<1$, or equivalently if $h>\eb^2$, $\dot{\phi}$ never
vanishes and the projection of $\mb{w}$ on the orbital plane of the binary moves along the full ellipse
\eqref{eq:hinterw}. In this case, $\mb{w}$ precesses around the angular momentum of the binary, $\mb{k}$. 
If $z>0$ the mutual inclination of the two orbits is always less than $90^\circ$ so the orbital motion of the
particle is prograde; conversely, if $z<0$ the mutual inclination of the two orbits is always superior to
$90^\circ$ so the orbital motion of the particle is retrograde.

If $k^2>1$ (or $h<\eb^2$), then $\dot{\phi}$ vanishes for $\phi_0=\pm \arcsin(1/k)$, changes its sign
(which is accompanied by a change of sign in the $z$ variable), and the angle $\phi$ librates between
$-\phi_0$ and $+\phi_0$. Thus, the projection of $\mb{w}$ on the orbital plane of the binary is bounded by
the unit circle to stay on an arc of ellipse \eqref{eq:hinterw}. In this case, $\mb{w}$ precesses around the
direction of perihelion of the binary, so that both the inclination and the node of the particle librate
around $\pm 90^\circ$.

In both cases, the period of the motion can be calculated with a simple quadrature using equation
\eqref{eq:dotphi}:

\begin{equation}\label{eq:Tpart}
 T=\frac{16}{3n_1}\frac{\MMb}{\nub}\left(\frac{a_2}{\ab}\right)^{7/2}
 \frac{K(k^2)(1-e_2^2)^2}{\sqrt{(1-\eb^2)(h+4\eb^2)}}\ ,\
\end{equation}
where $K(k^2)$ is the elliptic integral of the first kind defined by:
\begin{equation}
 K(k^2) = \left\{\begin{array}{cc}
                  \int_0^{\pi/2}\frac{d\phi}{\sqrt{1-k^2\sin^2\phi}} & \text{if}\ k^2<1\\
                  \int_0^{\phi_0}\frac{d\phi}{\sqrt{1-k^2\sin^2\phi}} & \text{if}\ k^2>1
                 \end{array}\right.\ .
\end{equation}

The last case where $k^2=1$ (or $h=\eb^2$) corresponds to the trajectories that separate the previous
two types. They link the points $(x,y,z)=(0,\pm 1,0)$, and the associated period is infinite. In the
projection on the $(x,y)$ plane, these separatrices form the ellipse which is tangent to the unit circle.
Since all trajectories that are inside this ellipse correspond to the precession of $\mb{w}$ around
$\mb{k}$, the width $\Delta x_{\text{sep}}$ of the separating ellipse in the $(x,y)$ plane gives an
indication on the proportion of such trajectories. Using equation \eqref{eq:xell} and the fact that 
$h=\eb^2$ on the separatrix, we get:

\begin{equation}
    \Delta x_{\text{sep}} = 2\sqrt{\frac{1-\eb^2}{1+4\eb^2}}
\end{equation}

Therefore, when the inner binary is circular, this width is equal to $2$, the full width of the unit circle,
and the only possible motion is precession of $\mb{w}$ around $\pm\mb{k}$. When the eccentricity of the binary
increases, the width of the separatrix decreases to zero, which is a limit case since it can only be reached
for a value of the binary's eccentricity equal to 1. The precession motions of $\mb{w}$ around $\pm\mb{i}$
thus become predominant when the eccentricity of the binary grows.

\subsection{Comparison with numerical studies}

\begin{figure}
 \includegraphics[width=\columnwidth]{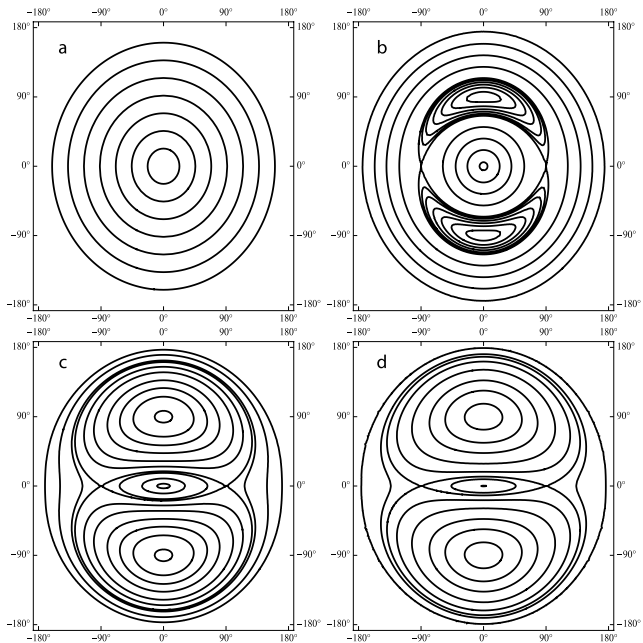}
 \caption{Energy levels of the Hamiltonian \eqref{eq:hmoy}
 in the $(i_2\cos\Omega_2,i_2\sin\Omega_2)$ plane for values of the eccentricity of the binary $\eb=0$ (a),
 $\eb=0.1$ (b), $\eb=0.79$ (c), $\eb=0.9$ (d).\label{fig:Heb}}
\end{figure}

In \citep{2009MNRAS.394.1721V}, the authors investigate a family of particles at high inclinations around the
binary HD 98800 Ba-Bb, which remain stable even under the perturbation of an outer third stellar companion.
They isolate a nodal precession imposed by the inner binary as the stabilizing mechanism working against the
destabilizing Kozai perturbations of the outer companion. They run simulations of test particles
orbiting the binary HD 98800 Ba-Bb using non secular equations. They observe the libration islands around
$i_2=\pm90^\circ$ and $\Omega_2=\pm90^\circ$ that we discussed in the previous section. As they show their results
in the $(i_2\cos\Omega_2,i_2\sin\Omega_2)$ plane, we plotted the energy levels of the outer restricted
Hamiltonian using these same coordinates for an easier comparison. Figure \ref{fig:Heb} shows these levels for
different values of the eccentricity. The c. panel in particular uses the
same value for the eccentricity of the binary ($\eb=0.79$) as figures 4 and 5 of \citep{2009MNRAS.394.1721V}.

Verrier and Evans notice no apparent structure in the dynamics of the $(e_2,\omega_2)$ couple apart from the
circulation of the perihelion. This is in agreement with the fact that the particle's eccentricity is
constant at the quadrupolar approximation. 

The authors also suggest that the projection of the angular momentum of test particles along the line of
apses of the binary may be an integral of motion. From the results of the previous section, it is
straightforward to see that the projection $x$ of the angular momentum of test particles along the line of
apses of the binary is not constant. It varies with an amplitude that decreases to $0$ when the inclination
of the particle approaches $\pm 90^\circ$, which can be misleading when looking at numerical results for
highly inclined particles. However, the norm of the angular momentum of the test particles is an integral of
the secular motion.

\begin{figure}
    \includegraphics[width=\columnwidth]{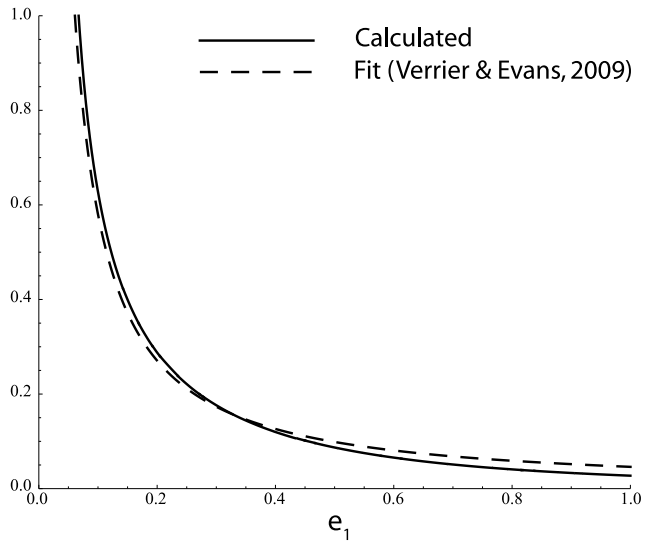}
    \caption{\label{fig:period} Dependence of the period \eqref{eq:Tpart} with respect to the eccentricity
    of the binary, in normalized units. The full line corresponds to the calculated period, while the dashed
    line corresponds to the power-law fit given by \citet{2009MNRAS.394.1721V}. We used a least squares method
    to fit the relative position of the two curves.}
\end{figure}

Finally, the authors give a power-law fit of the period of the libration of the node with respect to three
parameters: the eccentricity of the binary, the ratio of the semimajor axes $a_2/a_1$, and the mass ratio of
the binary, $\delta$. Their power-law is fitted using particles with fixed inclinations ($85^\circ$). They
give in their equation (5): 
\begin{equation} 
    T\propto e_1^{-1.1}\delta^{-0.8}\left(\frac{a_2}{a_1}\right)^{3.37}\ .
\end{equation}
 
By rewriting the mass dependences of equation \eqref{eq:Tpart}, we get the following analytical dependence
with respect to the mass ratio and the semi-major axis of the binary: 
\begin{equation}
    T\propto\left(\frac{a_2}{a_1}\right)^{3.5}(\delta(1-\delta))^{-1}\ .
\end{equation}

These two exponents compare very well with the fitted power law, in spite of the differences between the two
models. The dependency with respect to $e_1$ is rather complex in equation \eqref{eq:Tpart}, and it is best
compared in figure \ref{fig:period}.

The grid of initial conditions for the particles in \citep{2009MNRAS.394.1721V} extends however from 3 to 10
AU for a binary separation of 1 AU, so the quadrupolar approximation may not be sufficient to fully describe
the motion of the particles with the lowest semi major axes. In particular, Verrier and Evans state that some low
inclination particles show large eccentricity variations and even instability. This could be due to a low
initial semi-major axis and to resonances that are eliminated in our secular model by the averaging over the
mean anomalies.

\section{Problem of three massive bodies}\label{se:ftbp}

As we already stated, the quadrupolar secular three-body problem is still integrable when all the bodies have
positive masses. As such, it is possible to show how the outer restricted problem we discussed in the previous
section relates to the general case, and to the inner restricted case studied by \citet{1962AJ.....67..591K} and
\citet{1962P&SS....9..719L}. We will first express the Hamiltonian of the secular quadrupolar problem using the
same vectorial method as in the previous section in order to focus on the relative movements of the orbits. In
their studies of the secular quadrupolar problem, \citet{1976CeMec..13..471L} and \citet{1994CeMDA..58..245F}
have shown that this problem depends on two parameters. We will then point out which regions of parameter space
are topologically equivalent to the outer restricted case, and which regions correspond to the inner restricted
case, in order to show the continuity that exists between both situations.

\subsection{Hamiltonian} 

Let us consider three masses $m_0$, $m_1$ and $m_2$, this time with $m_2\neq0$.
We note the barycentric coordinates and impulsions $(\mb{u}_i,\mb{\tilde{u}}_i)_{i=0,1,2}$. As in the
previous section, we suppose that the two bodies of indices 0 and 1 form a binary and that the distance of
the third body to this binary is much larger than the separation of the binary. We still note $\delta =
m_0/(m_0+m_1)$. We first perform a canonical change of variables to Jacobi coordinates,

\begin{align}
 \mb{r}_0 &=\mb{u}_0 &\tilde{\mb{r}}_0 &= \tilde{\mb{u}}_0+\tilde{\mb{u}}_1+\tilde{\mb{u}}_2 =\mb{0} \\
 \mb{r}_1 &=\mb{u}_1-\mb{u}_0 &\tilde{\mb{r}}_1 &= \mb{\tilde{u}}_1 + (1-\delta)\tilde{\mb{u}}_2 \\
 \mb{r}_2 &=\mb{u}_2-\delta\mb{u}_0-(1-\delta)\mb{u}_1 & \tilde{\mb{r}}_2 &= \mb{\tilde{u}}_2
\end{align}

In these coordinates, the Hamiltonian of the three bodies is \citep{1989LaskarGoutelas}:
\begin{multline}
 H=\left(\frac{\tmb{r}_1^2}{2\beta_1}-\frac{\mu_1\beta_1}{r_1}\right) + \frac{\tmb{r}_2^2}{2\beta_2}\\
   -Gm_2\left(\frac{m_0}{|\mb{r}_2+(1-\delta)\mb{r}_1|}+\frac{m_1}{|\mb{r}_2-\delta\mb{r}_1|}\right),
\end{multline}

\noindent where $\beta_1^{-1}=m_0^{-1}+m_1^{-1}$, $\beta_2^{-1}=(m_0+m_1)^{-1}+m_2^{-1}$, $\mu_1=G(m_0+m_1)$
and $\mu_2=G(m_0+m_1+m_2)$.

Using the fact that $r_1\ll r_2$, we expand the Hamiltonian to order two in $r_1/r_2$ as in the previous
section:
\begin{multline}
 H=\left(\frac{\tmb{r}_1^2}{2\beta_1}-\frac{\mu_1\beta_1}{r_1}\right)+
 \left(\frac{\tmb{r}_2^2}{2\beta_2}-\frac{\mu_2\beta_2}{r_2}\right)\\
 -G\frac{\beta_1m_2}{2r_2^3}\left(3\frac{(\mb{r}_1.\mb{r}_2)^2}{r_2^2}-r_1^2\right)\ .
\end{multline}

The first two terms are Keplerian energies and are equal respectively to $-\mu_1\beta_1/2a_1$ and
$-\mu_2\beta_2/2a_2$, where $a_1$ and $a_2$ are the semi major axes of the inner and the outer body in our
system of coordinates.

We now average over the two mean anomalies $M_1$ and $M_2$ in order to get the secular part of the
Hamiltonian. We will define 4 unit vectors which are analogous to the 4 vectors we used in the first section:
$(\mb{i}_1,\mb{j}_1,\mb{k}_1)$ are tied to the orbit of the inner binary; $\mb{i}_1$ points in the direction
of the perihelion of the inner binary, $\mb{k}_1$ points in the direction of its angular momentum, and
$\mb{j}_1=\mb{k}_1\wedge\mb{i}_1$. The last vector $\mb{k}_2$ is colinear to the angular momentum of the
outer body. In this section, the vectors tied to the orbit of the inner binary will no longer have fixed
directions.

Using the same averaging formulae as in the previous section and using the fact that $(\mb{i}_1.\mb{k}_2)^2+
(\mb{j}_1.\mb{k}_2)^2+(\mb{k}_1.\mb{k}_2)^2=\mb{k}_2^2=1$, we can write:

\begin{multline}\label{eq:tmpavg2}
  \avg{H}_{M_1,M_2}=-\frac{\mu_1\beta_1}{2a_1}-\frac{\mu_2\beta_2}{2a_2}
  -\frac{3}{8}\frac{Gm_2\beta_1}{(1-e_2^2)^{3/2}}\frac{a_1^2}{a_2^3}\\
  \times\left[-\frac{1}{3}+2e_1^2+
  (1-e_1^2)\scal{\mb{k}_1}{\mb{k}_2}^2-5e_1^2\scal{\mb{i}_1}{\mb{k}_2}^2\right]\ .
\end{multline}

After averaging over the two mean anomalies, the semi-major axes are constant. There are thus four degrees of
freedom in the system, associated to the two eccentricities and the two inclinations. As we explained in the
previous section, the argument of perihelion of the outer body does not appear in the quadrupolar expansion,
and thus the norm of the angular momentum of the outer body $G_2=\beta_2\sqrt{\mu_2a_2(1-e_2^2)}$, is
constant. This implies that its eccentricity $e_2$ is constant. Using the reduction of the nodes would leave
only one degree of freedom in the reduced Hamiltonian, associated to the couple $(e_1,\omega_1)$. The full
reduction of the Hamiltonian and its expression in elliptical variables is the approach that has been used
widely, since it yields a very similar Hamiltonian function as in the inner restricted problem
\citep{1969CeMec...1..200H,1976CeMec..13..471L,1994CeMDA..58..245F}.

We want however to look at the motion of the nodes, or equivalently the motion of the vector $\mb{k}_2$ in
the moving frame $(\mb{i}_1,\mb{j}_1,\mb{k}_1)$ of the orbit of the second body.

In order to easily compute the equations of motion, we introduce two vectors associated to the orbit of the 
binary that are colinear to $\mb{i}_1$ and $\mb{k}_1$, and include in their norm the eccentricity of the 
binary, as in \citep{2009AJ....137.3706T}:
\begin{align}
 \mc{K}_1 &= \sqrt{1-e_1^2}\mb{k}_1\ , &\mc{I}_1 &=e_1 \mb{i}_1\ .
\end{align}

If we drop all the constant terms in equation \ref{eq:tmpavg2} and use the above vectors, we get:

\begin{equation}
 \avg{H}_{M_1,M_2}=-\frac{\alpha'G_2}{2}\left[2
 \mc{I}_1^2+(\mc{K}_1.\mb{k}_2)^2
                  -5(\mc{I}_1.\mb{k}_2)^2\right]\ ,
\end{equation}

\noindent where
\begin{equation}
 \alpha' = \frac{3}{4}n_1\left(\frac{a_1}{a_2}\right)^{7/2}\frac{\beta_1}{\MMb}\frac{1}{(1-e_2^2)^2}
        \sqrt{1+\frac{m_2}{\MMb}}\ ,
\end{equation}
and $\MMb,n_1$ are defined as in section \ref{se:rtbp}.

\subsection{Equations of motion}

The components of $\mc{K}_1,\mc{I}_1$ and $\mb{k}_2$ have the following Poisson brackets\footnote{We use the
following convention, where $p_i$ are momenta and $q_i$ positions: $\{f,g\}=\sum_i \left(\frac{\partial
f}{\partial p_i}\frac{\partial g}{\partial q_i} -\frac{\partial f}{\partial q_i}\frac{\partial g}{\partial
p_i}\right)$.} \citep{2005BorisovMamaev,2006Icar..185..312B}:

\begin{align}
 \{\mc{K}_{1i} ,\mc{K}_{1j} \} &= -\frac{\epsilon_{ijk}}{\Lambda_1} \mc{K}_{1k}\,, & 
           \{k_{2i} ,k_{2j} \} &= -\frac{\epsilon_{ijk}}{G_2} k_{2k}\,, \\
 \{\mc{I}_{1i} ,\mc{I}_{1j} \} &= -\frac{\epsilon_{ijk}}{\Lambda_1} \mc{K}_{1k}\,, & 
           \{\mc{K}_{1i} ,\mc{I}_{1j} \} &= -\frac{\epsilon_{ijk}}{\Lambda_1} \mc{I}_{1k}\,,
\end{align}
\noindent where $\Lambda_1=\beta_1\sqrt{\mu_1a_1}$ and $\epsilon_{ijk}$ is the Levi-Civita
symbol\footnote{$\epsilon_{ijk}=+1$ if $(i,j,k)$ is
an even permutation of $(1,2,3)$, $\epsilon_{ijk}=-1$ is the permutation is odd, and $\epsilon_{ijk}=0$ in
all other cases.}.

The equations of motion for the three vectors are thus:
\begin{eqnarray}
 \dot{\mc{K}}_1 & = & - \frac{1}{\Lambda_1} \left( \mc{K}_1 \wedge \nabla_{\mc{K}_1} H 
                      + \mc{I}_1 \wedge \nabla_{\mc{I}_1} H\right)\ , \\
 \dot{\mc{I}}_1 & = & - \frac{1}{\Lambda_1} \left( \mc{I}_1 \wedge \nabla_{\mc{K}_1} H 
                      + \mc{K}_1 \wedge \nabla_{\mc{I}_1} H\right)\ , \\
 \dot{\mb{k}}_2 & = & - \frac{1}{G_2} \mb{k}_2 \wedge \nabla_{\mb{k}_2} H\ .
\end{eqnarray}

In order to look at the motion of the vector $\mb{k}_2$ in the moving frame $(\mb{i}_1,\mb{j}_1,\mb{k}_1)$ of
the orbit of the second body, we use as \citet{2006Icar..185..312B} the above system to derive equations for
$x=(\mb{k}_2.\mb{i}_1)$, $y=(\mb{k}_2.\mb{j}_1)$, $z=(\mb{k}_2.\mb{k}_1)$ and $e_1$. Indeed,
$x=(\mb{k}_2.\mc{I}_1)/|\mc{I}_1|$, $z=(\mb{k}_2.\mc{K}_1)/|\mc{K}_1|$, $e_1=|\mc{I}_1|$, and $y$ is obtained
using the identity $x^2+y^2+z^2=1$:

\begin{align}
 \dot{x} & =  \alpha'(1-e_1^2)yz + \alpha'\frac{G_2}{\Lambda_1}\sqrt{1-e_1^2}y(2-5x^2) 
 \label{eq:fullmotion1}\\
 \dot{y} & =  -\alpha'(1+4e_1^2)xz  \nonumber \\
 &\phantom{=\dots}-\alpha'\frac{G_2}{\Lambda_1}\frac{x}{\sqrt{1-e_1^2}}[(1-e_1^2)(2-5x^2)+5e_1^2z^2] \\
 \dot{z} & =  5\alpha'e_1^2xy +\alpha'\frac{G_2}{\Lambda_1}\frac{5e_1^2}{\sqrt{1-e_1^2}}xyz \\
 \dot{e}_1 &=  \alpha'\frac{G_2}{\Lambda_1}5e_1\sqrt{1-e_1^2}xy \label{eq:fullmotion4}
\end{align}

The equations for $x,y$ and $z$ contain two terms: the first one is identical to the outer restricted system,
and the second one includes the motion of the reference frame $(\mb{i}_1,\mb{j}_1,\mb{k}_1)$ induced by the interaction
with the third body.
Note that when $\Lambda_1$ is very large compared to $G_2$ so that we can assume that $G_2/\Lambda_1$ is
equal to zero, which corresponds to the case where $m_2\ll m_0\text{ and }m_1$, the above system is identical to the
outer restricted system \eqref{eq:xdot}--\eqref{eq:zdot}.

\rev{The conservation of the total angular momentum $\mb{C}=\mb{G}_1+\mb{G}_2$ introduces the two main parameters
of the problem. Indeed,
\begin{equation}
 \Lambda_1^2(1-e_1^2) + G_2^2 + 2 \Lambda_1\sqrt{1-e_1^2} G_2 z = C^2\ .
\end{equation}

We note $\gamma=C/\Lambda_1$, $\gamma_2=G_2/\Lambda_1$.
The above expression of the norm of the total angular momentum can be rewritten as a second degree equation
giving $\sqrt{1-e_1^2}$ as a function of $z$ using the two parameters $\gamma$ and $\gamma_2$:
\begin{equation}\label{eq:redangmom}
 (1-e_1)^2 + 2\gamma_2z\sqrt{1-e_1^2} + \gamma_2^2-\gamma^2 = 0\ .
\end{equation}

The Hamiltonian can then be rewritten as:
\begin{equation}
 \avg{H}=-\frac{1}{2}\alpha'\Lambda_1\gamma_2[z^2+e_1^2(2-z^2-5x^2)]\ .
\end{equation}
}

The inequalities
$-1\leqslant z\leqslant 1$ and $0\leqslant e_1 < 1$ give the boundaries of the parameter space and the
range of possible values of $e_1$ for any given couple of parameters\footnote{The left part of the second
inequality is strict if $\gamma=\gamma_2$.} $(\gamma,\gamma_2)$: 
\begin{align} 
    |\gamma-\gamma_2|&\leqslant 1\ , \label{eq:paramspace}\\
    |\gamma-\gamma_2|&\leqslant \sqrt{1-e_1^2} \leqslant \min[\gamma+\gamma_2,1] \label{eq:allowedeta}\ . 
\end{align}

With these notations, the outer restricted problem of section \ref{se:rtbp} corresponds to the limit where
$\gamma_2=0$, and in this case $e_1=\sqrt{1-\gamma^2}$ is constant as we saw. Note that when
$\gamma_2>\gamma$, we have $G_2>C$, so this part of the parameter space contains only retrograde motions. Our
aim in this paper is to show the continuity between the outer restricted case we studied in section
\ref{se:rtbp}, and the inner restricted case that was investigated by \citet{1962AJ.....67..591K} and
\citet{1962P&SS....9..719L}. Both these problems lie in the region of parameter space where $\gamma>\gamma_2$
so we will restrict our study to this case\footnote{The other half of the parameter space ($\gamma\leqslant\gamma_2$)
corresponds to retrograde motions which are of less physical interest and much more technical to study using our
approach, in particular because equation \eqref{eq:redangmom} does not have a unique solution in this case. The interested
reader will find a complete discussion of this case in \citep{1976CeMec..13..471L,1994CeMDA..58..245F}.}.

%The above expression of the norm of the total angular momentum can be rewritten as a second degree equation
% giving $\sqrt{1-e_1^2}$ as a function of $z$:
% \begin{equation}\label{eq:redangmom}
%  (1-e_1)^2 + 2\gamma_2z\sqrt{1-e_1^2} + \gamma_2^2-\gamma^2 = 0\ .
% \end{equation}

In our case where $\gamma>\gamma_2$, there is only one acceptable root to \rev{equation \eqref{eq:redangmom}}, which is:
\begin{equation} 
    \sqrt{1-e^2_1}=-\gamma_2z+\sqrt{(\gamma_2z)^2+\gamma^2-\gamma_2^2}\ . 
\end{equation}

This relation implies that $e_1$ is a growing function of $z$. Note that $z=\cos i_2$, where $i_2$ is the inclination
of the outer body in the reference frame of the inner binary. As such, coplanar prograde motions ($z=1$) will always
occur for the maximal value of the eccentricity of the inner binary:
\begin{equation}
    e_{1,\max}=\sqrt{1-(\gamma-\gamma_2)^2}\ .
\end{equation}

Conversely, low eccentricities for the binary will be associated to lower values of $z$, and thus higher inclinations.
Relation \eqref{eq:allowedeta} implies that the inner binary can only have a circular motion if $\gamma+\gamma_2\geqslant 1$.
In this case, coplanar retrograde motion ($z=-1$) is not allowed, and the lowest value of $z$ is:
\begin{equation}\label{eq:z0}
   z_0 = \cos i_{2,\max} = \frac{\gamma^2-\gamma_2^2-1}{2\gamma_2}\ .
\end{equation}

When $\gamma+\gamma_2<1$ however, coplanar retrograde motion ($z=-1$) is possible and the associated value of the eccentricity
of the binary is:
\begin{equation}
 e_{1,\min}=\sqrt{1-(\gamma+\gamma_2)^2}\ .
\end{equation}

\begin{figure} 
	\includegraphics[width=\columnwidth]{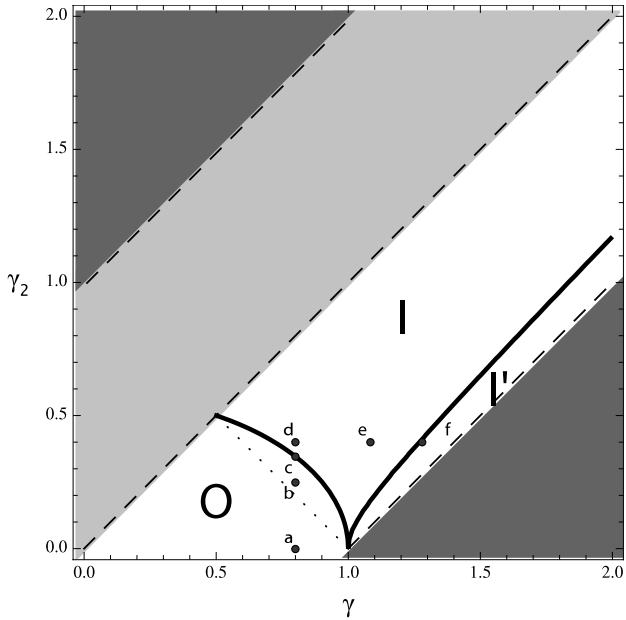}
	\caption{\label{fig:paramspace} Parameter Space. The dark gray areas are excluded by equation
\eqref{eq:paramspace}, the light gray area correspond to the part of parameter space corresponding to
$\gamma_2\geqslant\gamma$ which we do not study. The dotted line $\gamma+\gamma_2=1$ separates the zone where
there can be coplanar retrograde motion associated to a minimum eccentricity for the inner binary that is
strictly higher than $0$ (below the dotted line) and the zone where the inner binary can be circular but the
inclination is bounded (see section \ref{sse:eqcirc}). In zone O, the problem is topologically equivalent to the
outer restricted problem. In zones I and I' it is topologically equivalent to the inner restricted problem, with
zone I being equivalent to situations above the critical inclination and zone I' being equivalent to situations
under the critical inclination. The letters a--f correspond to the values of the parameters used to plot the
corresponding panels in figures \ref{fig:IOPlane} and \ref{fig:Spheres}.}
\end{figure}

\subsection{Dynamical regimes}\label{se:dynreg} 

\begin{figure} 
	\includegraphics[width=\columnwidth]{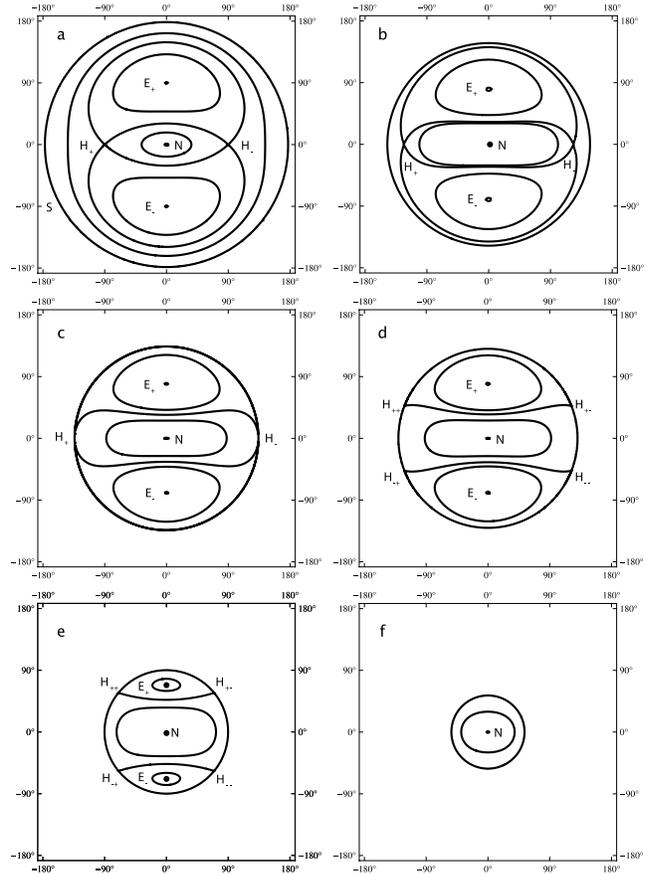} 
	\caption{\label{fig:IOPlane} Trajectories in the
$(i_2\cos\Omega_2,i_2\sin\Omega_2)$ plane for different values of the parameters. \rev{See section \ref{se:dynreg} for
a detailed discussion and appendix \ref{se:fixp} for calculations.} a: $(\gamma,\gamma_2)=(0.8,0)$
outer restricted case with $e_1=0.6$; b: $(\gamma,\gamma_2)=(0.8,0.25)$; c: $\gamma=0.8,\,
\gamma_2^2=(1/3)(1-\gamma^2)$; d: $(\gamma,\gamma_2)=(0.8,0.4)$; e: $(\gamma,\gamma_2)=(1.08,0.4)$; f:
$(\gamma,\gamma_2)=(1.28,0.4)$.} \end{figure}

\begin{figure} 
	\includegraphics[width=\columnwidth]{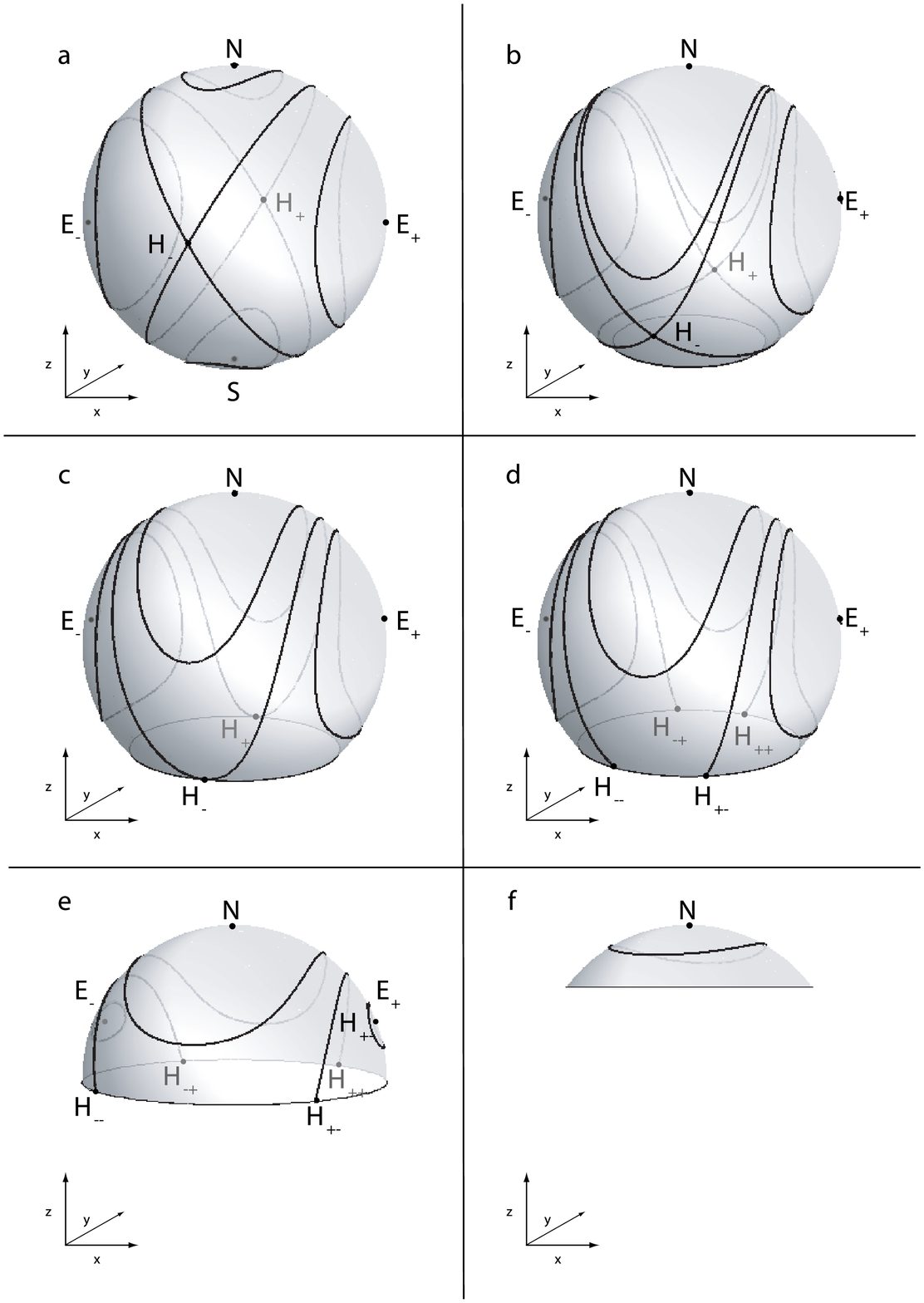} 
	\caption{\label{fig:Spheres} Trajectories on the
unit angular momentum sphere for different values of the parameters. \rev{See section \ref{se:dynreg} for
a detailed discussion and appendix \ref{se:fixp} for calculations.} a: $(\gamma,\gamma_2)=(0.8,0)$ outer
restricted case with $e_1=0.6$; b: $(\gamma,\gamma_2)=(0.8,0.25)$; c: $\gamma=0.8,\,
\gamma_2^2=(1/3)(1-\gamma^2)$; d: $(\gamma,\gamma_2)=(0.8,0.4)$; e: $(\gamma,\gamma_2)=(1.08,0.4)$; f:
$(\gamma,\gamma_2)=(1.28,0.4)$.} \end{figure}

In appendix \ref{se:fixp}, we briefly derive in the framework of the present study the fixed points of the
system and the boundaries of the dynamical regimes in parameter space that are given in
\citep{1994CeMDA..58..245F}. \rev{The fixed points are named as follow: the north pole is called $N$, and the south
pole $S$; linearly stable fixed points are named $E$, as elliptic, and linearly unstable points are named $H$, as
hyperbolic; finally, signs are placed as indices to refer to the symmetry of the problem with respect to the
two planes $x=0$ and $y=0$.} There are three dynamical regimes in the region of parameter space we study.

In region O of figure \ref{fig:paramspace}, the parameter $\gamma_2=G_2/\Lambda_1$ is small (less than 1/2).
This is the case in particular when the mass ratio $m_2/m_1$ is small. \rev{Moreover, $\gamma^2+3\gamma_2^2<1$.}
The phase space is topologically
equivalent to the outer restricted problem of section \ref{se:rtbp}. The north pole, which corresponds to
coplanar prograde motion with maximal eccentricity for the binary is linearly stable. There are two additional
stable fixed points $\text{E}_{\pm}$ in the plane $y=0$ \rev{(see section \ref{sse:eqy0})}.
They belong to the same family as the fixed points
$y=z=0$, $x=\pm 1$ of the outer restricted problem that are responsible for the stable high \rev{inclination} orbits
observed by \citet{2009MNRAS.394.1721V}. When $\gamma+\gamma_2\leqslant1$, the south pole which corresponds to
coplanar retrograde motion with minimal eccentricity for the binary, is also \rev{linearly} stable. \rev{The a panels
of figures \ref{fig:IOPlane} and \ref{fig:Spheres} provide a visualisation of the topology of this case.\footnote{\rev{Note that
the south pole in figure \ref{fig:IOPlane} a corresponds to the out-most trajectory; this is an artifact of the coordinate
map $(i_2\cos\Omega_2,i_2\sin\Omega_2)$ which sends the south pole of the sphere onto the circle $i_2=180^{\circ}$.}}}
When $\gamma+\gamma_2>1$, the south pole is no longer
accessible as stated in the previous section. It \rev{is however replaced by a stable trajectory} at a \rev{maximal}
inclination given by equation \eqref{eq:z0}, as can be seen on panel b \rev{of figures \ref{fig:IOPlane} and \ref{fig:Spheres}}.
This trajectory corresponds to a
circular inner binary \rev{(see section \ref{sse:eqcirc})}. Finally, there are two unstable points
$\text{H}_{\pm}$ in the $x=0$ plane that belong to
the same family as the unstable points of the outer restricted problem $x=z=0$, $y=\pm 1$ \rev{(see section \ref{sse:eqx0})}.

\rev{Panels c of figures \ref{fig:IOPlane} and \ref{fig:Spheres} show the limiting case between regions O and I. On this
boundary, $\gamma^2+3\gamma_2^2=1$. The two unstable points $H_{\pm}$ are now located on the boundary of the accessible
part of the sphere.}

\rev{Regions I and I' of figure \ref{fig:paramspace} are both in the part of the parameter space defined by
$\gamma^2+3\gamma_2^2>1$. In this zone, the problem becomes topologically equivalent to the inner restricted
problem studied by \citet{1962P&SS....9..719L} and \citet{1962AJ.....67..591K}. In the inner restricted case,
there is a critical value of the inclination ($\cos i_2=\sqrt{3/5}$) under which a circular inner binary is always
linearly stable, and above which a circular inner binary is always linearly unstable, giving rise to Kozai cycles.}

In region I of figure \ref{fig:paramspace}, the dynamical regime is topologically equivalent to the inner
restricted problem in the case where the
inclination is superior to the critical value. The limit trajectory $z=z_0$ which corresponds to a circular
inner binary becomes \rev{linearly} unstable. However, the north pole and the two fixed points $\text{E}_{\pm}$ are still
stable. In the inner restricted phase space, when the inclination is superior to the critical value, there are
two possible behaviours for the periastron of the inner particle: it can either circulate, or librate around
$\pm 90^{\circ}$. In our representation, the circulation case corresponds to trajectories around the north pole,
and the libration islands correspond to the two fixed points $\text{E}_{\pm}$. This is shown in panels d and e
in figures \ref{fig:IOPlane} and \ref{fig:Spheres}.

In region I' of figure \ref{fig:paramspace}, the dynamical regime is topologically equivalent to the inner
restricted problem in the case where the inclination is inferior to the critical value. Only one stable fixed
point remains, on the north pole of the sphere, associated to prograde coplanar motion. This is shown in panel f
in figures \ref{fig:IOPlane} and \ref{fig:Spheres}.

In both regions I and I', the parameter $\gamma_2=G_2/\Lambda_1$ can take higher values. This is in particular
true when the mass ratio $m_2/m_1$ increases.

The curve between regions I and I' is linked to the critical inclination that is defined in the inner
restricted case. Indeed, along that curve, given by equation \ref{eq:critlim}, we have the following limits when
$\gamma \rightarrow \infty$: 
\begin{align} 
	\frac{\gamma_2}{\gamma} &\rightarrow 1\, , & \gamma_2-\gamma \rightarrow -\sqrt{\frac{3}{5}}\ . 
\end{align}

When $G_2$ is very large compared to $G_1$, we can make the following first order expansion:

\begin{align}
 \gamma_2-\gamma &=\frac{G_2-C}{\Lambda_1} \\
  &=\frac{G_2-\sqrt{G_2^2+G_1^2+2(\mb{G}_2.\mb{G}_1)}}{\Lambda_1} \\
  &\approx \frac{G_2 - G_2(1+ (\mb{G}_2.\mb{G}_1)/G_2^2)}{\Lambda_1} \\
  &\approx - \frac{(\mb{G}_2.\mb{G}_1)}{\Lambda_1G_2} \\
  &\approx -z \sqrt{1-e_1^2} \ .
\end{align}

As such, we see that along the border between regions I and I', when $\gamma$ and $\gamma_2$ both tend to
infinity, we have the relation

\begin{equation}
 z \sqrt{1-e_1^2} \approx \sqrt{\frac{3}{5}}\ .
\end{equation}

Recall that $z=\cos i_2$, where $i_2$ is the inclination of the outer orbit in the reference frame of the 
inner orbit. Thus, the inclination of the inner orbit relatively to the outer orbit is $i_1=-i_2$, and the above
equation becomes:

\begin{equation}
 \cos i_1 \sqrt{1-e_1^2} \approx \sqrt{\frac{3}{5}}\ .
\end{equation}

This relation is precisely the one giving the critical value of the normal component of the angular momentum of
the inner body in the inner restricted problem.

\section*{Conclusion}

We first studied the case of a massless particle orbiting a binary at a long distance, and, in the secular and
quadrupolar approximations, gave a full analytical description of the motion along with the expression of the
period of the secular motion. When the inner binary is circular, only nodal precession takes place. However,
when the binary is elliptic, libration islands appear at high inclinations, and these islands grow bigger when
the eccentricity of the binary rises.
\citet{2008MNRAS.390.1377V,2009MNRAS.394.1721V} observe a similar nodal libration in their study of the stability
of particle populations in the quadruple stellar system HD 98800, and we showed that the analytical framework that
we derived for the outer restricted problem is well suited to explain the results of Verrier and Evans.

The quadrupolar secular three-body problem is still integrable when all the bodies have positive masses
\citep{1969CeMec...1..200H,1976CeMec..13..471L,1994CeMDA..58..245F}. Using a vectorial formalism as
\citep{2006Icar..185..312B,2009Icar..201..750B,2009AJ....137.3706T}, we looked at this problem from the point of
view of the outer restricted case. We showed how the outer restricted problem relates to the general case, and
to the inner restricted case studied by \citet{1962AJ.....67..591K} and \citet{1962P&SS....9..719L}: when the
mass of the outer body is small enough compared to the mass of the inner body, the general case behaves
similarly to the outer restricted problem. When the mass of the outer body increases enough, the general case
behaves like the inner restricted problem. We gave an expression of the boundary between these two regimes.

The outer restricted problem and its generalization to the non restricted case provide an interesting starting
point in the study of circumbinary planetary systems, such as the one discovered recently around the eclipsing
sdB+M system HW Virginis \citep{2009AJ....137.3181L}. In this system, the inner binary is very tight with a
period of 2.8 hr, while the proposed planetary companions have periods of 9.1 yr and 15.8 yr, so the quadrupolar
expansion is fully justified. Another field of application of the outer restricted problem is the study of the
motion of stars orbiting around binary black holes
\citep{2008AJ....135.2398M,2009ApJ...692.1075G,2009ApJ...693L..35M}.

\appendix

\section{Fixed points and bifurcations} \label{se:fixp}

The fixed points and boundaries presented in section \ref{se:dynreg} have already been studied by
\citet{1976CeMec..13..471L} and \citet{1994CeMDA..58..245F}. We briefly present here their derivation in the
framework of the present formalism. With the notations of section \ref{se:ftbp}, we will limit ourselves to
$\gamma>\gamma_2$.

\subsection{Poles of the sphere, $x=y=0$}\label{sse:eqpol}

This case corresponds to case 1 in section 5 of \citep{1994CeMDA..58..245F}. Note that their sphere is
constructed using the eccentricity and perihelion of the \rev{inner binary}, and is thus different from our angular
momentum sphere.

For all values of the parameters in the domain we study, the north pole $z=1$, which corresponds to coplanar
prograde motion, is a fixed point of the system. The associated eccentricity of the inner binary is:
\begin{equation}
    e_{1,\max}=\sqrt{1-(\gamma-\gamma_2)^2}\ .
\end{equation}
It is the maximal value of the eccentricity. This fixed point is always linearly stable. It is noted N in
figures \ref{fig:IOPlane} and \ref{fig:Spheres}. Figure \ref{fig:IOPlane} shows the lines of equal energy in
the $(i_2\cos\Omega_2,i_2\sin\Omega_2)$ plane, and figure \ref{fig:Spheres} shows these lines plotted on the
sphere of unit angular momentum of the outer body $\mb{k}_2^2=1$.

When $\gamma+\gamma_2<1$ (under the dotted line in figure \ref{fig:paramspace}), the south pole $z=-1$ (noted S
in the following figures), which corresponds to coplanar retrograde motion, is also a linearly stable fixed
point of the system. The eccentricity of the inner binary is minimal and equal to:
\begin{equation}
    e_{1,\min}=\sqrt{1-(\gamma+\gamma_2)^2}\ .
\end{equation}

Note that in this region of parameter space, the inner binary cannot be circular.

When $\gamma+\gamma_2\geqslant1$ (above the dotted line in figure \ref{fig:paramspace}), the minimal
eccentricity \rev{of} the binary is $0$ as deduced from \eqref{eq:allowedeta}. The south pole $z=-1$ does not
correspond to a real value of the eccentricity in this case. This limit however is not a bifurcation strictly
speaking. When crossing it, the stable south pole of the sphere \rev{is replaced by a stable trajectory at maximal
inclination}.

\subsection{Circular Trajectories for the inner binary}\label{sse:eqcirc}

In the region of parameter space where circular trajectories exist for the binary (above the dotted line in figure
\ref{fig:paramspace}), the value of $z$ which corresponds to such trajectories is minimal and equal to:
\begin{equation}
   z_0= \frac{\gamma^2-\gamma_2^2-1}{2\gamma_2}\ .
\end{equation}

The equations of motion on the small circle of the sphere $z=z_0$ are:
\begin{eqnarray}
 \dot{x}_{\phantom{1}} & = & \phantom{-} \alpha' y(z_0 + \gamma_2 (2-5x^2))\ , \label{eq:circmotionx}\\
 \dot{y}_{\phantom{1}} & = & - \alpha'x(z_0 +\gamma_2 (2-5x^2))\ , \label{eq:circmotiony}\\
 \dot{z}_{\phantom{1}} & = & \phantom{-} 0\ , \\
 \dot{e}_1 &= &\phantom{-}  0\ .
\end{eqnarray}

The right hand sides of equations \eqref{eq:circmotionx} and \eqref{eq:circmotiony} vanish for a certain
value of $x$ equal to:
\begin{equation}
    x_0^2 = \frac{\gamma^2+3\gamma_2^2-1}{10\gamma_2^2}\ .
\end{equation}

The curve $\gamma^2+3\gamma_2^2=1$ separates in figure \ref{fig:paramspace} the regions noted O and I. We can
distinguish three cases:
\begin{enumerate}
\item $\gamma^2+3\gamma_2^2<1$. In region O of figure \ref{fig:paramspace}, $x_0^2<0$ so there is no
fixed point on the circle $z=z_0$. As such, this circle is a trajectory for which the inner binary is circular
and the outer orbit precesses at a fixed inclination given by $i_{2,\max}=\text{ArcCos}\,z_0$. Moreover, this
trajectory is linearly stable.

\item $\gamma^2+3\gamma_2^2=1$. There are two fixed points on the circle $z=z_0$ at the coordinates
$(x=0,y=\pm\sqrt{1-z_0^2})$.

\item $\gamma^2+3\gamma_2^2> 1$. In this case, we must also check that $y_0^2=1-x_0^2-z_0^2\geqslant 0$.
The limit case where there is equality yields:
\begin{equation}
    5 \gamma_2^4 - (4 + 10 \gamma^2)\gamma_2^2 +  (5 \gamma^4  - 8 \gamma^2 + 3) = 0 \ .
\end{equation}
\end{enumerate}

This boundary limits the regions I and I' in figure \ref{fig:paramspace}. When solving the above equation for
$\gamma_2^2$ and selecting only the relevant solution satisfying $\gamma>\gamma_2\, , \gamma+\gamma_2\geqslant
1$, we obtain a solution that corresponds to equation 44 in section 5.2 of \citep{1994CeMDA..58..245F} and that
can be written using our notations as:
\begin{equation}\label{eq:critlim}
 \gamma_2^2 = \frac{2+5\gamma^2-\sqrt{60\gamma^2-11}}{5}\, , \quad \gamma+\gamma_2\geqslant 1\ .
\end{equation}

In region I, $y_0^2>0$ so there are four fixed points on the circle $z=z_0$, at the coordinates $(\pm x_0, \pm
y_0)$. They are noted $\text{H}_{\pm\pm}$, in panels d and e of figures \ref{fig:IOPlane} and \ref{fig:Spheres}.
Moreover, the trajectories that correspond to circular binaries are unstable in this zone. In region I' however,
$y_0^2<0$ so we are again in a region of parameter space where there are no fixed points on the circle $z=z_0$,
and the trajectories associated to circular binaries are again stable.

\subsection{The $x=0$ plane}\label{sse:eqx0}

When $x=0$, the only non trivial equation remaining in system \eqref{eq:fullmotion1}--\eqref{eq:fullmotion4}
is $\dot{x}=0$. Looking for a fixed point different from $x=y=0$, we have to solve
\begin{equation}
    z\sqrt{1-e_1^2}+2\gamma_2=0\ ,
\end{equation}

\noindent which after using relation \eqref{eq:redangmom} yields:

\begin{eqnarray} e_1 & = &\sqrt{1-\gamma^2-3\gamma_2^2}\ ,\\
x & = & 0 \ , \\
y & = & \pm\frac{\sqrt{\gamma^2-\gamma_2^2}}{\sqrt{\gamma^2+3\gamma_2^2}}\ ,\\
z & = & -\frac{2\gamma_2}{\sqrt{\gamma^2+3\gamma_2^2}}\ .
\end{eqnarray}

We thus have two symmetric fixed points in the plane $x=0$. They are noted $\text{H}_{\pm}$ in figures
\ref{fig:IOPlane} and \ref{fig:Spheres}. For these fixed points to exist, the associated
eccentricity must be a real number. As such, their domain of existence is the region noted O in figure
\ref{fig:paramspace}. This is case 2.1 in section 5 of \citep{1994CeMDA..58..245F}.

These two fixed points are linearly unstable in their domain of existence. Note that in the outer restricted
problem ($\gamma_2=0$), these fixed points are simply $x=z=0, y=\pm1$.

\subsection{The $y=0$ plane}\label{sse:eqy0}

When $y=0$, the only non trivial equation we must solve is $\dot{y}=0$. Here again, we look for another fixed
point than $x=y=0$, thus we have to solve:

\begin{equation}
 (1+4e_1^2)z +\frac{\gamma_2}{\sqrt{1-e_1^2}}[(1-e_1^2)(2-5x^2)+5e_1^2z^2]=0\ .
\end{equation}

Substituting $1-z^2$ in place of $x^2$ and then $\sqrt{1-e_1^2}$ in place of $z$ using \eqref{eq:redangmom},
we get:
\begin{equation}
 (1-e_1^2)^3 - \left(\gamma^2+\frac{1}{2}\gamma_2^2+\frac{5}{8}\right)(1-e_1^2)^2+\frac{5}{8}(\gamma^2-\gamma_2^2)^2=0
\end{equation}

This equation is the same as equation number 40 in \citep{1994CeMDA..58..245F}. In our
region of parameter space, there is at most one root which satisfies to the constraint \eqref{eq:allowedeta}.
The curve separating the zone where there is one solution and the zone where there is no solution corresponds
to the \rev{case where the limit value $e_1=0$ is a solution}, and coincides with the boundary between regions
I and I' in figure \ref{fig:paramspace} which is given by equation \eqref{eq:critlim}.

When there is a solution, the value of $e_1$ can be translated into a value of $z$ using
\eqref{eq:redangmom}. Since $y=0$, we get two values of $x=\pm \sqrt{1-z^2}$, and there are thus two
symmetric fixed points on the sphere, which are both linearly stable. They are noted $\text{E}_{\pm}$ in figures
\ref{fig:IOPlane} and \ref{fig:Spheres}. When $\gamma_2=0$, these fixed points
become simply $y=z=0$, $x=\pm 1$, which are responsible of the stable orbits at high inclination as discussed
in the previous sections.

\bibliography{FL2009}

\begin{thebibliography}{}

\bibitem[\protect\citeauthoryear{{Borisov} \& {Mamaev}}{{Borisov} \&
  {Mamaev}}{2005}]{2005BorisovMamaev}
{Borisov} A.~V.,  {Mamaev} I.~S.,  2005, {Dynamics of the Rigid Body} (in
  Russian).
R\&C Dynamics, Moscow, (http://ics.org.ru/)

\bibitem[\protect\citeauthoryear{{Bou{\'e}} \& {Laskar}}{{Bou{\'e}} \&
  {Laskar}}{2006}]{2006Icar..185..312B}
{Bou{\'e}} G.,  {Laskar} J.,  2006, Icarus, 185, 312

\bibitem[\protect\citeauthoryear{{Bou{\'e}} \& {Laskar}}{{Bou{\'e}} \&
  {Laskar}}{2009}]{2009Icar..201..750B}
{Bou{\'e}} G.,  {Laskar} J.,  2009, Icarus, 201, 750

\bibitem[\protect\citeauthoryear{{Ferrer} \& {Osacar}}{{Ferrer} \&
  {Osacar}}{1994}]{1994CeMDA..58..245F}
{Ferrer} S.,  {Osacar} C.,  1994, Celest. Mech. Dyn. Astron., 58, 245

\bibitem[\protect\citeauthoryear{{Gillessen}, {Eisenhauer}, {Trippe},
  {Alexander}, {Genzel}, {Martins} \& {Ott}}{{Gillessen}
  et~al.}{2009}]{2009ApJ...692.1075G}
{Gillessen} S.,  {Eisenhauer} F.,  {Trippe} S.,  {Alexander} T.,  {Genzel} R.,
  {Martins} F.,    {Ott} T.,  2009, ApJ, 692, 1075

\bibitem[\protect\citeauthoryear{{Harrington}}{{Harrington}}{1969}]{1969CeMec.%
..1..200H}
{Harrington} R.~S.,  1969, Celest. Mech., 1, 200

\bibitem[\protect\citeauthoryear{{Kinoshita} \& {Nakai}}{{Kinoshita} \&
  {Nakai}}{2007}]{2007CeMDA..98...67K}
{Kinoshita} H.,  {Nakai} H.,  2007, Celest. Mech. Dyn. Astron., 98, 67

\bibitem[\protect\citeauthoryear{{Kozai}}{{Kozai}}{1962}]{1962AJ.....67..591K}
{Kozai} Y.,  1962, AJ, 67, 591

\bibitem[\protect\citeauthoryear{{Laskar}}{{Laskar}}{1989}]{1989LaskarGoutelas}
{Laskar} J.,  1989, in {Benest} D.,  {Froeschle} C.,  eds, {Modern Methods in
  Celestial Mechanics,} {Systèmes de Variables et Eléments}.
Editions Frontières, pp 63--87

\bibitem[\protect\citeauthoryear{{Lee}, {Kim}, {Kim}, {Koch}, {Lee}, {Kim} \&
  {Park}}{{Lee} et~al.}{2009}]{2009AJ....137.3181L}
{Lee} J.~W.,  {Kim} S.-L.,  {Kim} C.-H.,  {Koch} R.~H.,  {Lee} C.-U.,  {Kim}
  H.-I.,    {Park} J.-H.,  2009, AJ, 137, 3181

\bibitem[\protect\citeauthoryear{{Lidov}}{{Lidov}}{1962}]{1962P&SS....9..719L}
{Lidov} M.~L.,  1962, P\&SS, 9, 719

\bibitem[\protect\citeauthoryear{{Lidov} \& {Ziglin}}{{Lidov} \&
  {Ziglin}}{1976}]{1976CeMec..13..471L}
{Lidov} M.~L.,  {Ziglin} S.~L.,  1976, Celest. Mech., 13, 471

\bibitem[\protect\citeauthoryear{{Malige}, {Robutel} \& {Laskar}}{{Malige}
  et~al.}{2002}]{2002CeMDA..84..283M}
{Malige} F.,  {Robutel} P.,    {Laskar} J.,  2002, Celest. Mech. Dyn. Astron.,
  84, 283

\bibitem[\protect\citeauthoryear{{Merritt}, {Gualandris} \&
  {Mikkola}}{{Merritt} et~al.}{2009}]{2009ApJ...693L..35M}
{Merritt} D.,  {Gualandris} A.,    {Mikkola} S.,  2009, ApJ Lett., 693, L35

\bibitem[\protect\citeauthoryear{{Mikkola} \& {Merritt}}{{Mikkola} \&
  {Merritt}}{2008}]{2008AJ....135.2398M}
{Mikkola} S.,  {Merritt} D.,  2008, AJ, 135, 2398

\bibitem[\protect\citeauthoryear{{Palaci{\'a}n} \& {Yanguas}}{{Palaci{\'a}n} \&
  {Yanguas}}{2006}]{2006CeMDA..95...81P}
{Palaci{\'a}n} J.~F.,  {Yanguas} P.,  2006, Celest. Mech. Dyn. Astron., 95, 81

\bibitem[\protect\citeauthoryear{{Palaci{\'a}n}, {Yanguas}, {Fern{\'a}ndez} \&
  {Nicotra}}{{Palaci{\'a}n} et~al.}{2006}]{2006PhyD..213...15P}
{Palaci{\'a}n} J.~F.,  {Yanguas} P.,  {Fern{\'a}ndez} S.,    {Nicotra} M.~A.,
  2006, Physica D Nonlinear Phenomena, 213, 15

\bibitem[\protect\citeauthoryear{{Poincaré}}{{Poincaré}}{1905}]{1905QB351.P73.%
.....}
{Poincaré} H.,  1905, {Leçons de mécanique céleste professées à la Sorbonne}

\bibitem[\protect\citeauthoryear{{Touma}, {Tremaine} \& {Kazandjian}}{{Touma}
  et~al.}{2009}]{2009MNRAS.394.1085T}
{Touma} J.~R.,  {Tremaine} S.,    {Kazandjian} M.~V.,  2009, MNRAS, 394, 1085

\bibitem[\protect\citeauthoryear{{Tremaine}, {Touma} \& {Namouni}}{{Tremaine}
  et~al.}{2009}]{2009AJ....137.3706T}
{Tremaine} S.,  {Touma} J.,    {Namouni} F.,  2009, AJ, 137, 3706

\bibitem[\protect\citeauthoryear{{Verrier} \& {Evans}}{{Verrier} \&
  {Evans}}{2008}]{2008MNRAS.390.1377V}
{Verrier} P.~E.,  {Evans} N.~W.,  2008, MNRAS, 390, 1377

\bibitem[\protect\citeauthoryear{{Verrier} \& {Evans}}{{Verrier} \&
  {Evans}}{2009}]{2009MNRAS.394.1721V}
{Verrier} P.~E.,  {Evans} N.~W.,  2009, MNRAS, 394, 1721

\end{thebibliography}

\end{document}